\documentclass{caps}
\usepackage{peter}
\usepackage{graphicx}

\begin{document}

\author[P. Hoeflich]{
P. H\"oflich, C. Gerardy, R. Quimby \\
University of Texas, Austin, TX 78712, USA}   
\chapter{Asphericity Effects  in Supernovae}
\begin{abstract}
 We present a brief summary of asphericity effects in thermonuclear
and core collapse supernovae (SN), and how to distinguish the underlying
physics by their observable signatures. Electron scattering is the dominant process to
produce polarization which is one of the main diagnostical tools. Asphericities result in
a directional dependence of the luminosity which has direct implications for
the use of SNe in cosmology.
For core collapse SNe, the current observations and their interpretations 
suggest that the explosion mechanism itself is highly aspherical with a well defined axis and, typically,
axis ratios of 2 to 3.
 Asymmetric density/chemical distributions and off-center energy depositions
have been identified as crucial for the interpretation of the polarization $P$.
 For thermonuclear SNe, polarization turned out to be an order of magnitude smaller strongly
supporting rather spherical, radially stratified envelopes. Nevertheless, asymmetries have
been recognized as important signatures  to
probe A) for the signatures of the progenitor system, B) the
global asymmetry with well defined axis,
 likely to be caused by rotation of an accreting white dwarf or merging WDs,
 and C) possible remains of the deflagration pattern. 
\end{abstract}

\section{Introduction}
During the last decade, advances in observational, theoretical and 
computational astronomy have provided new insights into the nature and 
physics of SNe and gamma-ray bursts. Due to the extreme brightness of
these events, they are expected to continue to play important role in 
cosmology. SNe~Ia allowed good measurements of the Hubble constant both 
by statistical methods and theoretical models. SNe~Ia have provided one 
of the strong evidences for a non-zero cosmological constant, and will 
be used to probe the nature of the dark energy. In the future, core 
collapse SNe and the related GRBs may become the tool of choice to probe the very
first generation of stars.  The multidimensional nature of
these objects has been realized and it has become obvious that 
measurements of asymmetries and their understanding is a key for
understanding of both SNe and GRBs. In this contribution, we want
to give a brief overview on the observable consequences
 of asphericity including a directional dependence of the luminosity with
a special emphasis on polarization in Thomson scattering dominated envelopes.
In a first part, we want to give a brief introduction to the configurations which
produce polarization. Subsequently, we present examples for various mechanisms
which produce polarization in core collapse and thermonuclear SNe.
 It is beyond the scope to present a review of the current literature.
A more general
overview about the physics core collapse SNe with references to the general literature can be found in
H\"oflich et al. (2002). For thermonuclear SNe on scenarios and details of the nuclear burning
front, we want to refer  H\"oflich et al. (2003), and Hillebrandt \& Niemeyer (2000), Khokhlov (2001)
and Gamezo \& Khokhlov (this volume), respectively.
W will focus on the theoretical aspects. For a complementary discussion
of the observations, see Wang(this volume).
 
\begin{figure}[hb]
\vskip -0.5cm
\includegraphics[width=6.cm,angle=270,clip=]{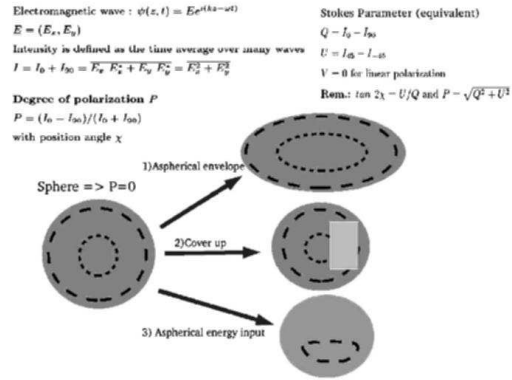}
\vskip -0.5cm
\caption {
 Definition of the polarization and schematic diagram for its production.
 The dotted lines give the
main orientation of the electrical vectors.
 For an unresolved sphere, the components
cancel  out (from H\"oflich 1995).
}
\vskip -0.7cm
\label{cases}
\end{figure}

\section{General}
\begin{figure}[ht]
\vskip -0.1cm
\includegraphics[width=3.0cm,angle=270,clip=]{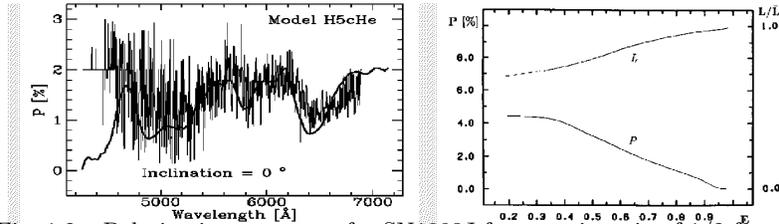}
\vskip -0.5cm
\caption {
Polarization spectrum for SN1993J for an axis ratio of 1/2 for an oblate ellipsoid
in comparison with observations by
Trammell et al. (1993) (left plot). On the right,
the dependence of the continuum polarization (right) and directional
dependence of the luminosity is shown 
 as a function
axis ratios for oblate  ellipsoids  seen from the equator
(from H\"oflich, 1991 \& H\"oflich et al. 1995).
}
\vskip -0.6cm
\label{93j}
\end{figure}

Asymmetry can be probed by direct imaging of ejecta of the remnants, e.g. in SN1987A
(Wang et al. 2002, H\"oflich et al. 2001b), or Cas A (Fesen \& Gunderson 1997), 
proper motions of neutron stars (Strom et al. 1995),
or, more general,  by polarization measurements during the early phase of the expansion.
 In SN, polarization is mainly produced by Thomson scattering of photons in
an aspherical configuration. It can be caused by asymmetries in the density, abundances or 
excitation structure of an envelope.
In general, the  ejecta cannot be spatially resolved. Although the light from 
different parts
of a spherical disk is polarized, the resulting polarization $\bar P$ is zero for 
the integrated light (Fig. \ref{cases}).
 To produce $\bar P$, three basic configurations may  be considered, in which I) the photosphere
 is aspherical, II) parts of
 the disk are  shaded, and III) the envelope may be illuminated by an off-center light source.
 In case II, the shading may be either by a broad-band  absorber such as dust or a specific line opacity.
In the latter case, this would produce a change of $\bar P$ in a narrow line range (Figs. \ref{93j}).
 In reality, a combination of all cases may be realized.
Note that quantitative analyzes of SNe need to take into
account that the continua and lines are not formed in the same layers.
 Polarization carries the information about the apparent, global asymmetries. E.g., it increases with increasing
axis ratios in ellipsoidal geometries or off-center energy sources but decreases due to multiple
scattering (the optical depth) or steep density profiles. Observed size of $P$ depends also on the
position of the observer relative to the object. For a detailed discussion see H\"oflich
(1991, 1995). As a consequence, the interpretation of polarization data are not unique.
 Another problem is due to polarization by the interstellar medium.
 In parts, these limitations can be overcome by spectropolarimetry and time series of observations which
provides additional information due to spectral features  (see Fig. \ref{93j}) and their evolution.
Still, the use of consistent physical models is mandatory to further constrain the variety of interpretation.

\section{Core Collapse Supernovae}
In recent years, there has been a mounting evidence that the explosions of massive stars (core
collapse SNe) are highly aspherical.
 The spectra (e.g., SN87A, SN93J, SN94I, SN99em, SN02ap)
 are significantly polarized at a level of 0.5 to 3 \% 
  (M\'endez et al. 1988, Cropper et al. 1988, H\"oflich 1991, Jeffrey 1991)
 indicating aspherical envelopes by factors of up to 2 (see Fig. \ref{93j}).
The degree of polarization  tends to vary inversely with the mass of the hydrogen
envelope, being maximum for Type Ib/c events with no hydrogen 
(Wang et al. 2001). For SNeII, Leonard et al. (2000) and Wang et al. (2001) showed  that the
polarization and, thus, the asphericity increase  with time.
The orientation of the polarization vector tends to stay constant
both in time and with wavelength.  This implies that there is a  global symmetry
axis in the ejecta.  Both trends suggest a connection
of the asymmetries with the central engine which may be understood in terms of jet-induced
explosions (Khokhlov et al. 1999, H\"oflich et al. 1998, 2002), or pulsational modes in neutrino driven
explosions (Scheck et al. 2003).
\begin{figure}[ht]
\vskip -0.3cm
\includegraphics[width=6.1cm,angle=270,clip=]{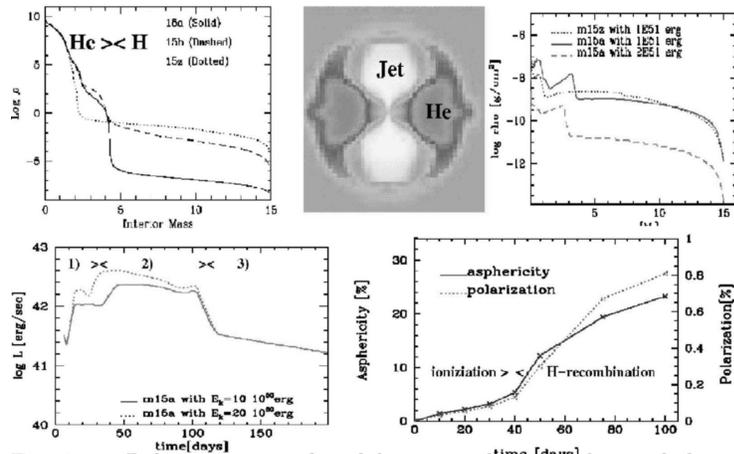}
\vskip -0.5cm
\caption{ Polarization produced by an aspherical, chemical distribution
for an extreme SN~IIp model such as SN1999em (see text).
}
\vskip -0.5cm
\label{99em}
\end{figure}

 However,  even strongly asymmetric explosions do not produce  asymmetries in the
massive hydrogen-rich envelopes of SNeII which are sufficiently large
to explain the polarization observed in SN1987A or SN1999em.
Aspherical excitation by hard radiation  is found to be crucial.
 As example, the extreme 
 SNIIp 1999em is shown in Fig. \ref{99em} and, for details, see H\"oflich et al. (2002).
Our calculations of the initial stage of the explosion employ 3D hydrodynamics.
 The explosion of a star with 15 solar masses is triggered by a low velocity,
 high density jet/bipolar outflow which delivers a explosion energies of
 of 1 and  $2 \times 10^{51}erg$. The jet from the central engine  stalls after about 250 seconds, and
the abundance distribution freezes out in the expanding envelope.
  The resulting distribution of the the He-rich layers is given in Fig. \ref{99em}.
The colors white, yellow, green, blue and red correspond to He mass fractions of
0., 0.18, 0.36, 0.72, and  1., respectively. The composition of the jet-region consists of
a mixture of heavy elements with about 0.07 $M_\odot$ of radioactive $^{56}Ni$. After about 100 seconds,
the expansion of the envelope becomes spherical. Thus, for times larger than 250 seconds,
the explosion has been followed in 1-D up to the phase of homologous
expansion. In the upper, right panel, the density distribution is given at about 5 days
after the explosion.  The steep gradients in the density in the upper right and left panels
are located at the interface between the He-core and the H-rich mantel.
 In the lower, left panel,
the resulting bolometric LCs are given for explosion energies of
 2E51erg (dotted line) and 1E51erg, respectively.
 Based on full 3-D calculations for the radiation \& $\gamma $-ray transport,
we have calculated the location of the recombination front (in NLTE) as a function of
time. The resulting shape of the photosphere is always prolate.
 The corresponding axis ratio and the  polarization seen from the equator are shown 
 (lower, right panel).
Note the strong increase of the asphericity after the onset of the recombination phase between
day 30 to 40 (H\"oflich et al. 2002). For the polarization in a massive, H-rich envelope,
$P$  seems to be directly liked to the recombination process and asymmetric excitation.

\section{Thermonuclear Explosions}
\begin{figure}
\vskip -0.1cm
\includegraphics[width=5.0cm,angle=270]{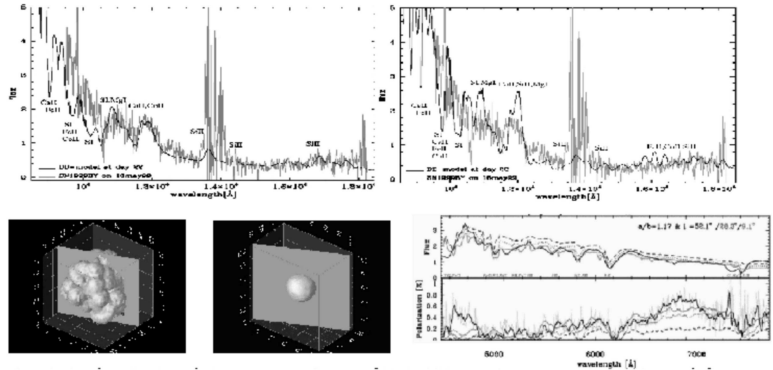}
\caption{ Analysis of the subluminous SN1999by using a combination of flux and polarization
data. {\bf Upper panel:}
Comparison of the  NIR spectrum on
May 16 (left) with a spherical, subluminous delayed detonation model.
 For this object, the spectra are formed in layers of explosive carbon and incomplete silicon
burning up to about 2 weeks after maximum light. This is in strict contrast to normal bright SNe~Ia
where the photosphere enters layers of complete Si burning already at about maximum light.
 On the right, we show a comparison of the observed and theoretical
spectrum if we impose mixing of the inner 0.7 $M_\odot$ as
it can be expected based on detailed 3-D deflagration models (Khokhlov, 2001).
Strong homogeneous mixing of the inner layers can be ruled out because the excess
excitation of intermediate mass elements and the absorption by iron-group elements(from H\"oflich et al. 2002).
{\bf Lower, left panel:} Energy deposition by $\gamma $-rays at day 1 (left) and 23 (right)
based on our full 3-D MC gamma ray transport based on a $^{56}Ni$
distribution of a typical deflagration models.
 The diameter of the WD is normalized to 100. At about day 23, the energy deposition
is not confined to the radioactive $^{56}Ni$ ruling out clumpiness as a solution
to the problem mentioned above (from H\"oflich 2002).
{\bf Lower, right panel:} Optical flux and polarization spectra
 at day 15 after the explosion for the subluminous 3-D delayed-detonation model in comparison with the
SN1999by at about  maximum light. The interstellar component of $P$ has been determined
to $P=0.25 \% $  with a polarization angle of $140^o$.
 The observed flux and the smoothed polarization spectra are the solid black lines.
The light grey line is  the original data for P at a resolution of 12.5 \AA .
In the observations, the polarization angle is constant indicating  rotational symmetry of the envelope.
 The structure of the spherical model has been mapped into oblate ellipsoids with axis ratios A/B
of 1.17  (from Howell, H\"oflich, Wang \& Wheeler 2001).}
\vskip -0.4cm
\label{99by}[t]
\end{figure}

 For thermonuclear explosions, polarization turned out to be an important tool to
probe for
A)  the  global asymmetry caused by the WD, i.e. the continuum component,
B) for the signatures of the progenitor system,
 and C) possible remains of the deflagration pattern produced during the early phase of nuclear 
 burning.

Case A:
 In general, the maximum, continuum polarization is about an order of magnitude
smaller than in core collapse SNe and it is decreasing with time but, again, with a
well defined axis of symmetry (e.g. Wang et al. 2003). This decrease occurs
despite a significant contribution of electron scattering to the  opacity 
till about 1 to 2 weeks after the explosion.
 Overall, the objects are rather spherical as could be expected for thermonuclear
explosions of a WD. For the subluminous SN1999by,  the
continuum was polarized up to about 0.7 \% (Howell et al. 2001).
The well defined axis can be understood in the framework of
rotating WDs which may be a consequence the accretion process in a binary system
or the merging of two WDs.
 For the strongly subluminous SN 1999by, the accretion
on an accreting WD is clearly favored from the analysis of light curves and spectra (Fig. \ref{99by}).
 Its unusually high continuum polarization may suggest a correlation between the
subluminosity and its low Ni production, i.e. the propagation of the nuclear burning front.

\begin{figure}
\includegraphics[width=3.8cm,angle=270]{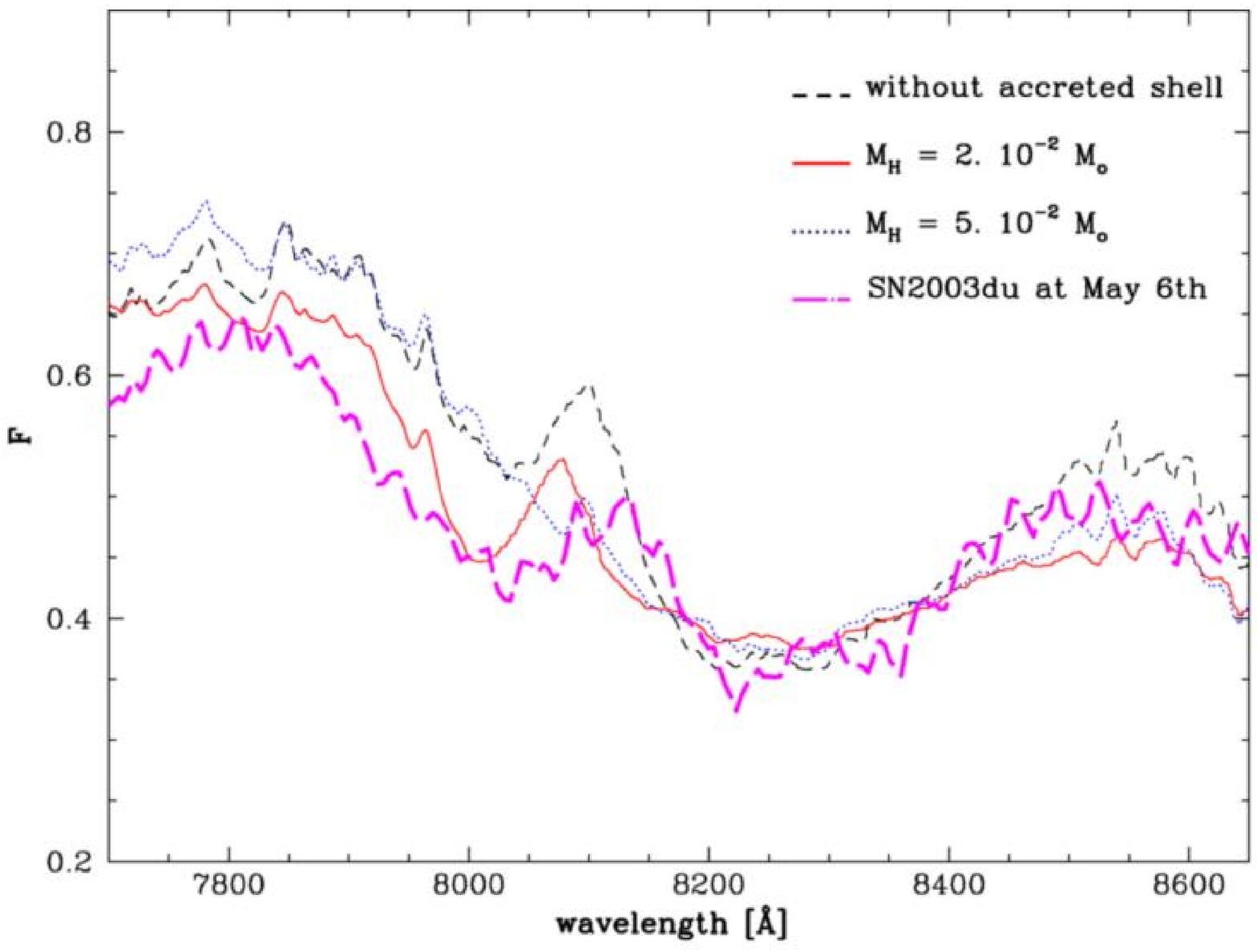}
\includegraphics[width=4.0cm,angle=270]{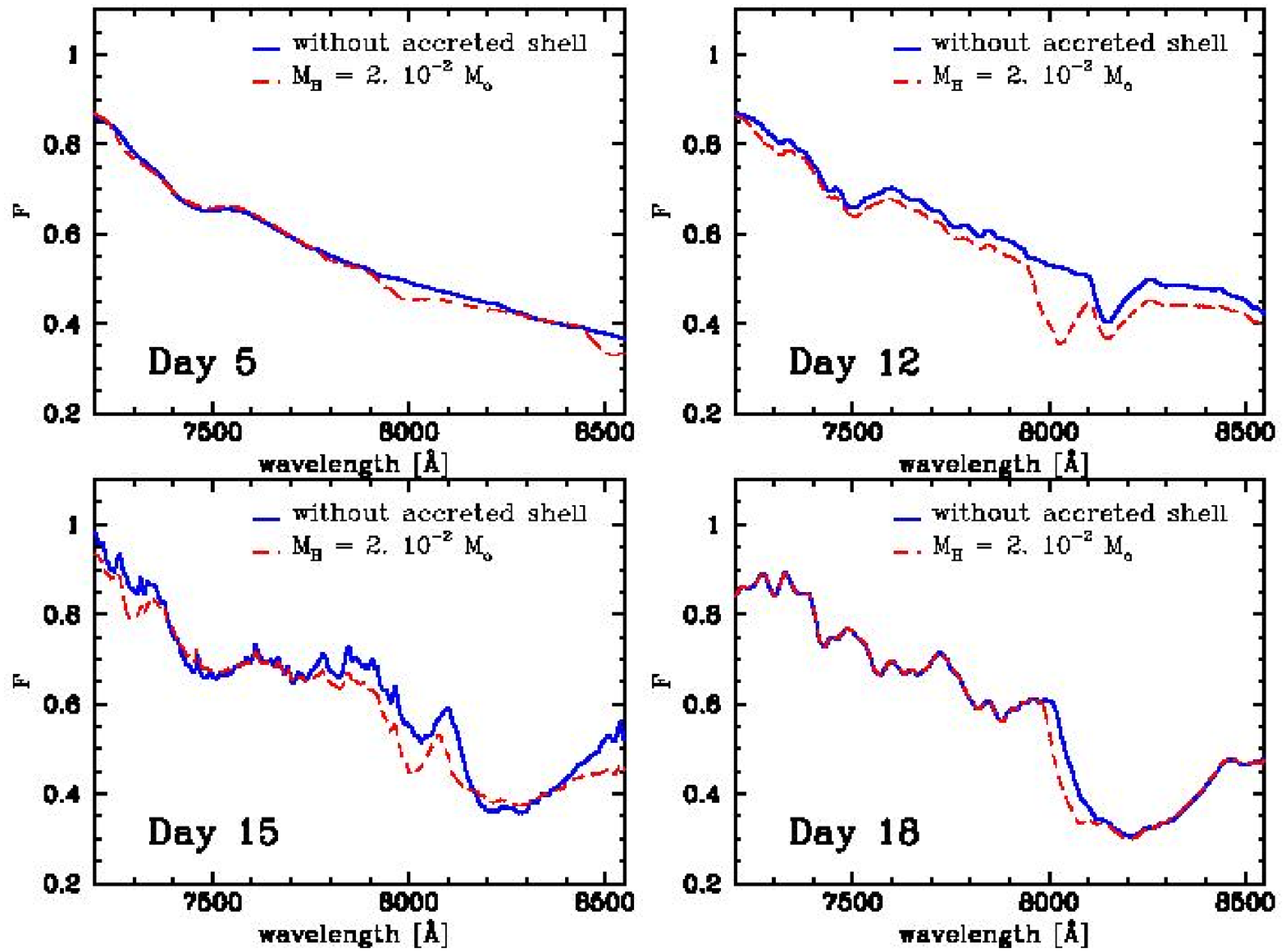}
\caption{CaII IR feature observed in SN~2003du on May 6 in comparison with theoretical models
at about 15 days after the explosion, and its evolution with time.
 The models are based on a delayed detonation model which
interacted  with a H-rich shell of 0.02 and 0.05 $M_\odot$ during the early phase of the explosion.
 The dominant signature of this interaction is the appearance of a secondary, high velocity Ca II feature or,
for high shell  masses, a persistent high velocity component in a broad Ca~II line.
Without ongoing interaction, no H or He lines are detectable.
 Note that, even without a shell, a secondary Ca II feature can be seen for a period of 2 to 3 days
during the phase when Ca III recombines to Ca II emphasizing the importance of a good time coverage
for the observations.}
\vskip -0.7cm
\label{03du}
\end{figure}

Case B: Interaction with the circumstellar environment may be detected by the appearance of a high
velocity component in Ca II (Fig. \ref{03du}).  In SN1994D, it may be understood as an ionization
effect when Ca III recombines to CaII and, thus, forming two, radially separated features
(H\"oflich et al. 1998, Hatano et al. 2001). Alternatively, double features of CaII  may 
be attributed to abundance pattern (Fisher et al. 1997). The observed polarization in SN2001el
was high in Ca~II clearly identifying this feature as a morphological distinct pattern but
its origin remained open (Wang et al. 2003, Kasen et al. 2003).
 Based on a detailed analysis of SN2003du, Gerardy et al. (2003) showed that this feature can be understood
in the framework of the interaction between the SN eject and its H-rich nearby surroundings.

Case C: Wang et al. (1998) showed that the observed polarization pattern may consistent with chemical
inhomogeneities at the Si/Ni interface as can be expected from 3-D deflagration models. However,
this detection was on a $1 \sigma $ level, and needs to be confirmed in other objects.


\begin{thereferences}{99}
\bibitem{} Cropper M., Bailey J., McCowage J., Cannon R., Couch W. 1988, MNRAS 231, 685
\bibitem{}Fesen, R. A. \& Gunderson, K. S. 1996, ApJ, 470, 967
\bibitem{}Fisher A., Branch D.,  Nugent P., Baron E. 1997, 481L, 89
\bibitem{gerardy2000} Gerardy C., H\"oflich P., Quimby R., Wang L. + the HET-SN team 2003, ApJ, submitted
\bibitem{hatano2000} Hatano K., Branch D., Lentz E.J., Baron E.,
Filippenko A. V., Garnavich P. 2000, ApJ  543, L94
\bibitem{hn00} Hillebrandt, W., Niemeyer, J. 2000, ARAA 38, 191
\bibitem{}H\"oflich, P. 1991 A\&A 246, 481
\bibitem{}H\"oflich, P. 1995, { ApJ} {443}, 89
\bibitem{}H\"oflich P.,  Wheeler, J.C., Hines, D., Trammell S. 1995, ApJ 459, 307
\bibitem{}H\"oflich, P., Khokhlov A., Wang L., 2002, AIP-Publ.  586, p. 459 \& astro-ph/0104025
\bibitem{hetal03} H\"oflich, P., Gerardy, C., Linder, E., \& Marion, H.
2003, in: Stellar Candles, eds. Gieren et al., Lecture Notes in Physics, Springer Press,
in press \& astro-ph/0301334
\bibitem{howell01} Howell A., H\"oflich P., Wang L., Wheeler J. C. 2001,
ApJ 556, 302
\bibitem{}Jeffrey  D.J., 1991, ApJ, 375, 264
\bibitem{kasen2003} Kasen, D., Nugent, P., Wang, L., Howell, A., Wheeler, J. C.,
H\"oflich, P., Baade, D., Baron, E., Hauschildt, P. 2003, ApJ , in press \& astro-ph/0301312
\bibitem{khokhlov01} Khokhlov, A. 2001, astro-ph/0008463
\bibitem{}Leonard D.C., Filippenko, A.V., Barth A.J., Matheson T.  2000, ApJ 536, 239
\bibitem{}Mendez R.H. et al. 1977,  ApJ 334, 295
\bibitem{}Scheck L., Plewa T., Janka H.-T., Kifonidis K., M\"uller E. 2003, Phys.Rev.Let, submitted
\bibitem{}Strom R., Johnston H.M., Verbunt F., Aschenbach B. 1995, Nature, 373, 587
\bibitem{}Trammell S., Hines D., Wheeler J.C. 1993, ApJ 414, 21
\bibitem{}Wang L., Wheeler J.C., Li Z., Clocchiatti A., 1996, ApJ 467, 435
\bibitem{}Wang L., Howell A., H\"oflich P., Wheeler C. 2001, ApJ 550, 1030
\bibitem{}Wang L., Baade D., H\"oflich P., Wheeler C., Fransson C., Lundqvist P. 2002, ApJ, submitted 
\bibitem{}Wang L., et al.  2002b, ApJ Let., in press \& astro-ph/0205337 
\bibitem{wang2003a}  Wang, L., Baade, D., H\"oflich, P., Khokhlov, A, Wheeler,
J. C., Kasen,  D., Nugent  P., Perlmutter S., Fransson C., Lundqvist P. 2003, ApJ 591, 1110
\bibitem{wang2003b} Wang, L., Baade, D., H\"oflich, P., Wheeler, J. C.,
Kawabata, K., Nomoto, K. 2003, ApJ, submitted
\end{thereferences}
\end{document}